\documentclass[aps,prl,preprint,groupedaddress,showpacs]{revtex4}

\usepackage{graphicx}
\usepackage{amsmath}
\usepackage{epsfig}

\begin{document}

\title{High-brilliance synchrotron radiation induced by the plasma magnetostatic mode}

\author{F. Fiuza}
\email[Electronic address: frederico.fiuza@ist.utl.pt]{}

\author{L. O. Silva}
\email[Electronic address: luis.silva@ist.utl.pt]{}
\affiliation{GoLP/Instituto de Plasmas e Fus\~ao Nuclear - Laborat\'orio Associado, Instituto Superior T\'ecnico, Lisboa, Portugal}

\author{C. Joshi}
\affiliation{Department of Electrical Engineering, University of California, Los Angeles, California 90095, USA}

\date{\today}

\begin{abstract}
Using multi-dimensional PIC simulations we show that the magnetic undulator-type field of the plasma magnetostatic mode is indeed produced by the interaction of a laser pulse with a relativistic ionization front, as predicted by linear theory for a cold plasma. When the front with this magnetostatic mode is followed by a relativistic electron beam, the interaction of the beam with this magnetic field, produces FEL-type synchrotron radiation, providing a direct signature of the magnetostatic mode. The possibility of generating readily detectable ultrashort wavelength radiation using this mode, by employing state-of-the-art laser systems, is demonstrated, thus opening the way towards experimental observation of the hitherto unseen magnetostatic mode and the use of this plasma FEL mechanism to provide a source of high-brilliance ultrashort wavelength radiation.

\end{abstract}



\maketitle

The advent of high power lasers has allowed for the exploration of the extraordinary ability of plasmas to sustain strong electric \cite{bib:joshi2} and magnetic \cite{bib:tatarakis} fields. Electrostatic plasma waves \cite{bib:langmuir}, driven by intense laser pulses have well-defined electric structures that can be used to accelerate electron beams up to the GeV energy \cite{bib:dawson,bib:dreambeam,bib:leemans,bib:tsung}. Laser-plasma interactions are also known to provide ultra-strong ($\gtrsim$ 100 T) magnetic fields \cite{bib:tatarakis,bib:bfield}. However, effective methods for generating controlled periodic magnetostatic structures in plasmas are yet to be demonstrated.

In a recent study, it was shown that an overdense ionization front moving nearly at $c$ can frequency upshift microwave radiation confined in a cavity  \cite{bib:bmodeexp}. This study, and earlier theoretical work \cite{bib:bmodeth}, has led to the understanding of how electromagnetic boundary conditions should be handled at a plane dielectric discontinuity moving normal to the plane. A zero-frequency, zero-electric-field, purely Amp\`ere-law magnetostatic mode (here called the MS mode) was recognized as being important. This mode has hitherto not been seen and is ordinarily very difficult to excite. With a plasma front boundary which is moving at a normal velocity $v_{f}$, the general phase continuity condition is that the quantity ($\omega - k \cdot v_{f}$) does not change when changing from one medium to the other in $\omega-k$ space \cite{bib:bmodeth}. With a fast-moving front and an incident EM (ElectroMagnetic) wave whose frequency $\omega_i$ is well below the plasma frequency ($\omega_p$) behind the front, the stationary MS mode becomes easy to set up, and the MS mode (which has the same magnetic polarization as the incident EM wave) becomes the dominant repository for the incident energy \cite{bib:bmodeth}. (The value of $\omega_p$ is usually that obtained after rapid and complete ionization by a high frequency $(\omega_d^2 \gg \omega_p^2)$ ionization front driver laser pulse of the gas behind the ionizing pulse.) The MS mode will have a normal wavevector $k_{MS}$ of $(1+v_{f}/c)$ times the incident wavevector (in the case of interest this factor is very nearly 2). Since the MS mode does not propagate and cannot exist outside the plasma, it has not yet been detected directly.

For a plane polarized incident EM wave the MS mode in the plasma has the magnetic field geometry of standard magnetic undulators, such as those used for FELs (Free Electron Lasers), and from this arises the key concept of this letter in addressing the MS mode detection problem. One begins by creating a usefully strong MS undulator-type B-field using the sub-plasma-frequency laser wave incident on the counter-propagating ionization front created by a short ionization-front laser drive pulse incident from the opposite side of the plasma. The key new component here is to supply also a highly relativistic electron beam pulse to pass through the stationary MS undulator magnetic field almost immediately after it is produced. This is done in order to produce a pulse of synchrotron radiation more or less in the electron beam direction, thereby confirming the existence of the MS mode.

This MS synchrotron radiation will provide both a direct diagnostic of the MS spatial period (from the frequency spectrum peak) and a reasonable estimate of its strength (from the magnitude of the synchrotron signal). Going beyond this basic verification of the MS mode physics brings in the complementary notion of using the concept to produce a powerful undulator magnetic field of much shorter spatial period than is possible to obtain with conventional magnets. Applications might range from condensed matter physics \cite{bib:barut,bib:ye} to novel radiations sources \cite{bib:fel}. Showing how this would work in practice is the purpose of the rest of this paper, as exemplified by the simulation results and the further detailed discussions.

The efficient generation of the MS mode requires a sharp ionization front, i.e. $L_g k_{i} \ll 1$, where $L_g$ is the gradient scale of the ionization front and $k_i$ is the wavenumber of the incident light \cite{bib:bmodeth}. Steep ionization fronts can be generated by the propagation of a short intense laser pulse in a gas jet via tunneling ionization. The relativistic factor of a laser-driven ionization front, in the linear regime, is given by $\gamma_f = \omega_d/\omega_p$. For plasmas transparent to the ionizing light, $\omega_d \gg \omega_p$, and for $\omega_i^2 \gg \omega_p^2/4 \gamma_f^2$ the amplitude of the MS mode \cite{bib:bmodeth} can be written as
\begin{equation} 
B_{MS} [\mathrm{T}] \simeq \frac{58}{1+0.8\left(\frac{5\times10^{19}}{n_e [\mathrm{cm}^{-3}]}\right) \left(\frac{10.6}{\lambda_i [\mu \mathrm{m}]}\right)^2} \left(\frac{I_i [\mathrm{W}/\mathrm{cm}^2]}{10^{13}}\right)^{1/2},
\label{eq:bmode}
\end{equation} 
where $\lambda_i = 2 \pi c/\omega_i$ is the wavelength of the incident radiation and $I_i$ its intensity, which is only limited by the threshold for ionization of the background gas. For Hydrogen, for instance, the threshold intensity for tunnel ionization \cite{bib:adk} is $\sim 10^{13}$ W$/$cm$^2$ for mid-infrared light, allowing for the generation of magnetic-field amplitudes on the order of 100 T.

\begin{figure}
\begin{center}
\includegraphics[width=0.6\textwidth]{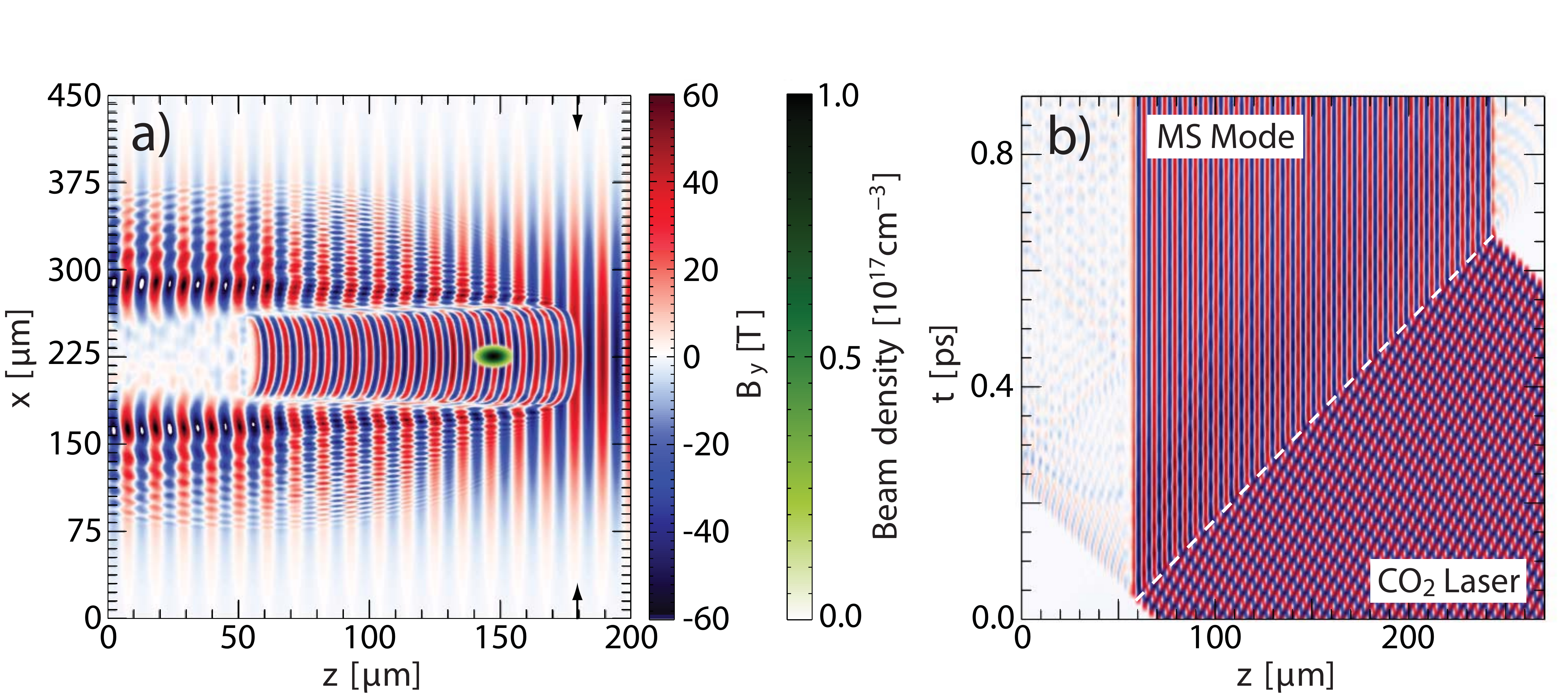}
\caption{\label{fig:beam_bmode}(color online) a) Snapshot of the MS mode generated by a CO$_2$ laser pulse ($\lambda_i = 10.6 ~\mu$m, $\epsilon_i = 8$ mJ, $W_i = 165~\mu$m, $\tau_i = 1.25$ ps, $I_i \simeq 3 \times 10^{13}$W/cm$^2$) in 2D PIC simulations. A short pulse ($\lambda_d = 0.4 ~\mu$m, $\epsilon_d = 2$ mJ, $W_d = 50~\mu$m, $I_d \simeq 2 \times 10^{15}$W/cm$^2$, $\tau_d = 50$ fs) is launched from the left hand side of the simulation box through an Hydrogen gas jet ($n_0 = 5\times10^{19}$ cm$^{-3}$) and is trailed by an e-beam (energy $\epsilon_b = 100$ MeV, charge $Q_b = 10$ pC, duration $\tau_b = 25$ fs, radius $r_b = 5 ~\mu$m, and density $n_b = 10^{17}$ cm$^{-3}$). The CO$_2$ laser pulse is launched from the right hand side of the simulation box. The simulation box size is $450\times450$ $\mu$m$^2$, with $2^{15}\times2^{11}$ cells and 9 particles per cell (roughly 600 million particles were used). b) Time streak of a longitudinal lineout of the magnetic field at the center. The arrows and the dashed line indicate the position of the front.}
\end{center}
\end{figure}

\begin{figure}
\begin{center}
\includegraphics[height=0.3\textwidth]{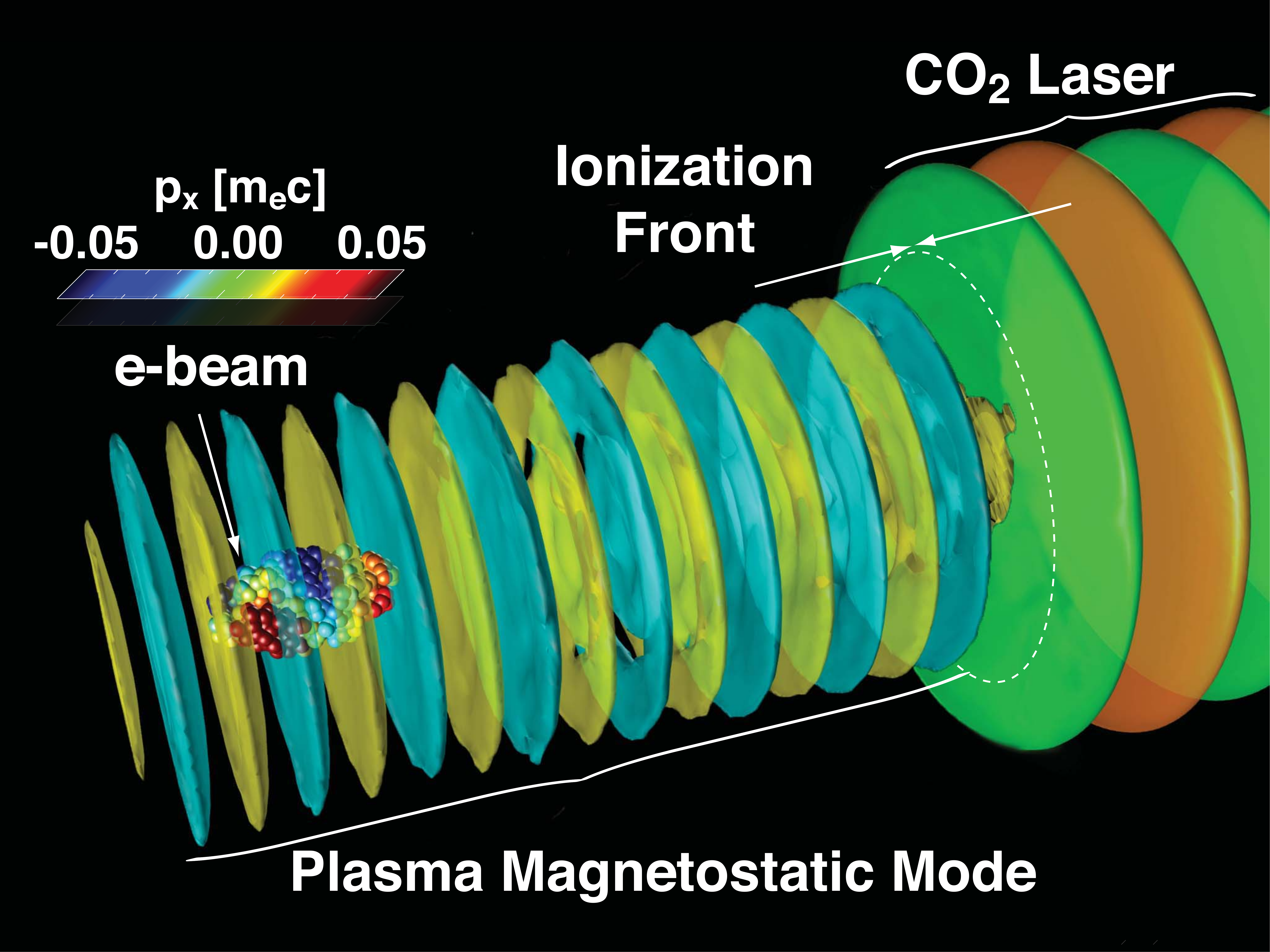}
\caption{\label{fig:beam_bmode2}(color online) 3D snapshot of the MS mode. Beam electrons are colored by transverse momentum and the B-field of the CO$_2$ laser is truncated at the ionization front position. The laser and e-beam parameters are identical to those in Fig. \ref{fig:beam_bmode} for a scaled simulation box with a width of 180 $\mu$m and a length of 450 $\mu$m.}
\end{center}
\end{figure}

Present state-of-the-art CO$_2$ laser systems \cite{bib:tochitsky} can deliver pulses with energy $\epsilon_i = 170$ J and $\lambda_i = 10.6 ~\mu$m, thus allowing for the generation of a MS amplitude of 100 T, a wavelength of 5.3 $\mu$m, and a controllable width and length (e.g. with spot size of $400 ~ \mu$m and length $\sim 10$ cm), with the MS mode-width determined by the smaller of the spot size of the incident pulse, $W_i$, and the width of the ionization front (approximately the spot size of the driver pulse, $W_d$), and the MS length is typically determined by the incident pulse duration, $\tau_i$. Figs. \ref{fig:beam_bmode} and \ref{fig:beam_bmode2} illustrate the features of the MS mode generated by the collision of a CO$_2$ laser pulse with a relativistic ionization front in two-dimensional (2D) and three-dimensional (3D) OSIRIS 2.0 \cite{bib:fonseca} PIC simulations. The ionization front is self-consistently generated by tunnel ionization due to a short, linearly polarized laser pulse. The CO$_2$ laser pulse is launched in counter-propagation with the ionization front. We have also launched an electron beam trailing the driver pulse in order to probe the generated MS mode. We can observe that behind the ionization front a modulated magnetic field is induced, with a wavelength of 5.3 $\mu$m and amplitudes of 54 T, for $n_0 = 5\times10^{19}$ cm$^{-3}$ (Fig. \ref{fig:beam_bmode}), and 93 T, for $n_0 = 5\times10^{20}$ cm$^{-3}$ (Fig. \ref{fig:beam_bmode2}), in very good agreement with Eq. \eqref{eq:bmode}. Moreover, there is almost no transmission of the CO$_2$ pulse in that region. The MS mode (with wavevector $2 k_i$) is stationary in space behind the ionization front, which itself moves to the right at $\sim c$, while the incident field (with wavevector $k_i$) moves to the left at $c$ (Fig. \ref{fig:beam_bmode}b). The transverse momentum of the beam electrons, $p_x$, shown in Fig. \ref{fig:beam_bmode2}, illustrating the wiggling motion in the MS mode, is in very good agreement with the theoretical prediction $|p_x|_{max} = K$ \cite{bib:jackson}, where $K = 0.934 B_u [\mathrm{T}] \lambda_u [\mathrm{cm}]$ is the dimensionless undulator strength parameter, $B_u$ is the amplitude of the undulator magnetic field, and $\lambda_u$ is the wavelength of the undulator. The presence of the trailing e-beam does not affect the MS structure, indicating that the MS mode can be used as a magnetic undulator.

As the beam electrons wiggle in the undulator structure they emit radiation with wavelength $\lambda_r = \lambda_u (1 + K^2/2)/2\gamma_b^2$ \cite{bib:jackson}, where $\gamma_b$ is the relativistic factor of the beam electrons. The short wavelength ($\lambda_u \sim 5 ~ \mu \mathrm{m}$) and high amplitude ($B_u \sim 100 ~\mathrm{T}$) laser driven MS structure allows for the generation of radiation with features that can provide an evidence of the MS mode (energy determined by the properties of the e-beam and MS structure, narrow spectrum, and high brilliance). The synchrotron radiation emitted by the e-beam traversing the MS mode, and therefore the MS mode itself, can be unambiguously detected through a straightforward null test whereby the e-beam/incident laser pulse are present or not in the experimental set-up.

The total power radiated by an e-beam traversing the MS undulator-like field is given by $Q_b P/e$, where $P = (1/3)e^4 B_u^2 \gamma_b^2 /m_e^2 c^3$ \cite{bib:jackson} is the power radiated by a single electron in the undulator structure, yielding $P_r [\mathrm{kW}]$$\simeq$$0.71 \left(B_u [\mathrm{T}]/60\right)^2 \left(\epsilon_b [\mathrm{MeV}]/100\right)^2 \left(Q_b [\mathrm{nC}]/0.1\right)$. The average radiated energy is $P_r \tau_p/2$, where $\tau_p$ is the propagation time of the e-beam in the MS structure. Since the residence time of the e-beam in the MS structure is half the duration of the incident wave, $\tau_i$, the total radiated energy can be written as
\begin{equation}
\epsilon_r [\mathrm{nJ}] \simeq 28.49 \left(\frac{B_u [\mathrm{T}]}{60}\right)^2 \left(\frac{\epsilon_b [\mathrm{MeV}]}{100}\right)^2 \frac{Q_b [\mathrm{nC}]}{0.1} \frac{\tau_i [\mathrm{ps}]}{160}.\label{eq:energy}
\end{equation} 
The energy of the radiated photons is
\begin{equation} 
\epsilon_\gamma [\mathrm{keV}] \simeq 18.72 \frac{10.6}{\lambda_i [\mu\mathrm{m}]} \left(\frac{\epsilon_b [\mathrm{MeV}]}{100}\right)^2,
\label{eq:ph_energy}
\end{equation}
which depends only on the energy of the e-beam and on the MS wavelength. We note that while the MS magnetic field can be strong, $K$ is still small due to the short wavelenght of the MS undulator, and, therefore, the photon energy is well approximated by $4 \gamma_b^2$ times the incident laser photon energy. The total number of photons, $N_\gamma=\epsilon_r/\epsilon_\gamma$, is given by $N_\gamma [10^7] \simeq 0.95 (B_u [\mathrm{T}]/60)^2 (\lambda_i [\mu\mathrm{m}]/10.6) (Q_b [\mathrm{nC}]/0.1) (\tau_i [\mathrm{ps}]/160)$. For typical parameters, that can be reached in an experiment, $10^7$ photons are emitted in a half cone angle of $1/\gamma_b$, and can be detected far away from the plasma, using an x-ray CCD camera in a single photon mode (where as few as 3-5 photons can be detected in the energy range of interest) and suitable spectral filtering. It is also important to point out that the e-beam propagates along a large number of periods, $N_u$, in the undulator-like structure. Thus, for a low energy spread e-beam, $\Delta \epsilon_b/\epsilon_b \ll 1$, the energy spread of the emitted radiation, $\Delta \epsilon_\gamma/\epsilon_\gamma$, is narrow, since it is determined by the largest of $1/N_u$ or $\Delta \epsilon_b/\epsilon_b$. The dimensionless peak brilliance $\hat{B}_\gamma$ of the radiated photon beam, normalized to 1 photon/s/mrad$^2$/mm$^2$/0.1$\%$ bandwidth, is $\hat{B}_\gamma [10^{28}] \simeq 1.52 (B_u [\mathrm{T}]/60)^2$$(\tau_i [\mathrm{ps}]/160)^2$$(3/r_b [\mu \mathrm{m}])^2$ $(0.1/\theta [\mathrm{mrad}])^2$$(Q_b [\mathrm{nC}]/0.1)$$(10/\tau_b [\mathrm{fs}])$, where $r_b$ is the radius of the e-beam and $\theta$ its angular divergence.  The propagation of an e-beam with $\epsilon_b = 1$ GeV through the plasma undulator induced by a CO$_2$ laser pulse, would provide gamma-ray pulses with $\epsilon_\gamma \sim$ 2 MeV and a peak brilliance ranging from $\hat{B}_\gamma \sim 5 \times 10^{25}$, for typical laser wakefield accelerator (LWFA) e-beams ($Q_b = 30$ pC, $\theta = 1$ mrad, $\tau_b = 10$ fs, $r_b = 3 ~\mu$m), to $\hat{B}_\gamma \sim 10^{30}$ for e-beams from typical conventional linear accelerators ($Q_b = 2$ nC, $\theta = 1 ~\mu$rad, $\tau_b  = 100$ fs, $r_b = 60 ~\mu$m).

A proof-of-principle experiment, demonstrating and measuring the MS mode can be performed using the widely available 1 J class Ti:Sapphire laser systems, whose laser pulses can efficiently generate both the relativistic ionization front and LWFA e-beams with $\epsilon_b = 50$ MeV and $Q_b = 30$ pC \cite{bib:dreambeam}. Using an improved two-stage Raman shifting scheme with $>0.2\%$ efficiency \cite{bib:wada}, a fraction of the laser energy (0.2 J) can be used to produce pulses with $\tau_i =  4$ ps and $\lambda_i = 5 ~\mu$m, which then collide with the ionization front at incident intensities $I_i \simeq 3 \times 10^{13}$ W/cm$^2$, $W_i = 10 ~\mu$m, inducing the MS mode. With these parameters, $N_\gamma \simeq 10^5$, peaked at $\epsilon_\gamma \simeq 10$ keV, which would allow for a clear evidence of the MS mode.

Since the e-beam probing the MS mode propagates in a plasma, it is important to assess the possible beam quality degradation, in terms of energy spread and emittance, due to beam-plasma and laser-plasma instabilities. The relevant beam-plasma instabilities are the two-stream instability (longitudinal) and the Weibel instability (transverse). The two-stream instability can be avoided using short e-beams, $L_b$[mm] $\lesssim$ ($\epsilon_b$[MeV]/100) ($n_b$[$10^{17}$cm$^{-3}$])$^{-1/3}$ ($n_e$[$5\times10^{19}$cm$^{-3}$])$^{-1/6}$ \cite{bib:joshi}, where $n_b$ is the electron density of the beam. The Weibel instability is strongly suppressed for e-beams narrower than the plasma wavelength $\lambda_p = 2 \pi c/\omega_p$ \cite{bib:katsouleas}, or for wider low-density e-beams with $n_b/n_e < 2 \times 10^{-3}(\epsilon_b [\mathrm{MeV}]/100) (\theta [\mathrm{\mu rad}]/10)^2$ \cite{bib:weibel}. In what concerns laser-plasma instabilities, laser filamentation is avoided for laser powers smaller than the critical power for self-focusing \cite{bib:focusing}, and stimulated Raman scattering \cite{bib:raman} will not be driven for ultrashort laser pulses, which are required in order to generate a steep ionization front. Laser wakefields \cite{bib:dawson,bib:wakes}, which can also have deleterious effects on the e-beam quality, are negligible for a laser normalized vector potential $a_{0,d} \ll 1$, since $\delta n_e/n_{e0} \ll 1$.

Ionization front irregularities can also contribute to a degradation of the MS mode and consequently of the emitted radiation. Erosion of the head of the laser pulse can broaden the initially sharp ionization front and, therefore, the laser should contain enough energy to generate a stable ionization front during the interaction process. The energy depletion length, $L_{dpl}$, due to ionization can be roughly estimated by equating the laser energy to the ionization energy of a single atom, $W_{ion}$, times the number of atoms ionized, $n_0 \sigma_d \tau_d$ (where $\sigma_d$ is the laser cross section), yielding $L_{dpl}$[cm]$ \simeq 10 I_{d} [2 \times 10^{16} $Wcm$^{-2}] \tau_{d}[50 $fs$] / (n_0 [5 \times 10^{19}$cm$^{-3}] W_{ion}[13.6$eV$])$. For typical experimental parameters, $L_{dpl}$ exceeds several centimeters. Phase irregularities across the ionization front can also lead to a local variation of the MS wavelength, increasing the energy spread of the emitted radiation. However, near the axis this energy spread induced by phase irregularities should be very small (the energy spread is proportional to $1-$cos$[\phi]$, where $\phi$ is the variation in angle of the tilted wavefront), and therefore, for e-beams that are narrower than the ionization front (driver pulse) this should not significantly affect the brilliance of the emitted radiation. 2D simulations of the ionization front propagation in lengths in excess of 1 cm have confirmed this, and will be presented elsewhere.

\begin{figure}
\begin{center}
\includegraphics[height=0.3\textwidth]{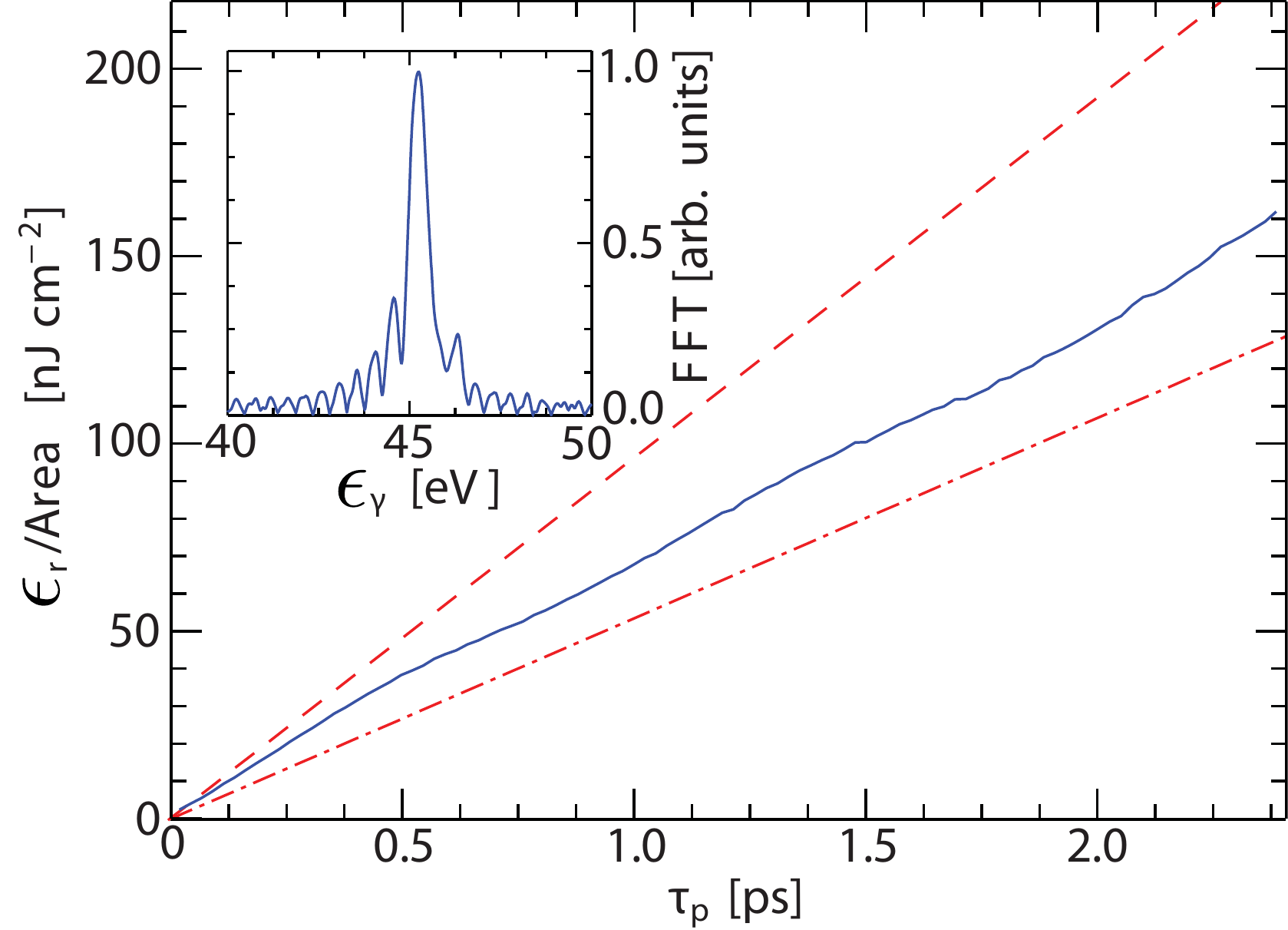}
\caption{\label{fig:radiation}(color online) Energy radiated by the propagation of an e-beam ($\epsilon_b = 5$ MeV, $\tau_b = 10.5$ fs, $n_b = 8.2 \times 10^{16}$ cm$^{-3}$) through the MS mode (solid), generated by the collision of a CO$_2$ laser pulse with a relativistic ionization front, in 1D PIC simulations. The laser parameters (ionization-front driver and incident CO$_2$ laser) are identical to those in Fig. \ref{fig:beam_bmode} and the Hydrogen gas jet has a neutral density $n_0 = 5\times10^{20}$ cm$^{-3}$. The simulation box is $1.5$ mm long, with $2^{19}$ cells and 100 particles per cell (total $\sim$ 52 million particles). The theoretical radiated energy, Eq. \eqref{eq:energy}, is plotted for the initial (dashed) and final (dot-dashed) B-field amplitude. After 1.8 ps of propagation ($N_u = 100$), the radiation spectrum exhibits a narrow energy spread $\Delta\epsilon_\gamma/\epsilon_\gamma = 0.012$ (shown in the inset).}
\end{center}
\end{figure}

We have performed 1D PIC simulations fully resolving the radiation generated due to the interaction of the relativistic e-beam with the induced MS mode, with a numerical subtraction technique to isolate the synchrotron radiation of the e-beam; by running two identical simulations, with and without the e-beam, the radiated fields are obtained from the subtraction of the results of the two simulations. Fig. \ref{fig:radiation} illustrates the temporal evolution of the energy radiated by the e-beam, showing the generation of radiation with $\epsilon_\gamma = 45.2$ eV. The induced MS amplitude varies from 93 T (when the e-beam starts to interact with the structure) to 69 T (at the end of the interaction), due to the slow increase of the rise time of the ionization front as the driver pulse propagates \cite{bib:fiuza}. The total radiated energy and the radiation spectrum are in good agreement with Eqs. \eqref{eq:energy} and \eqref{eq:ph_energy}.

In conclusion, we have proposed a configuration that allows for the detection and characterization of the hitherto unseen plasma MS mode. It is induced by the collision of light pulses with relativistic ionization fronts, and detected by using it as a plasma undulator. PIC simulations show the generation of periodic, short wavelength magnetic fields associated with the excitation of the MS mode, and illustrate the generation of ultrashort wavelength synchrotron radiation by the propagation of an electron beam through this magnetic structure. Our results open the way towards the first experimental evidence of the MS mode from the measurements and analysis of the emitted radiation. Once this is accomplished, the possibility of using the plasma MS mode with an electron beam to produce an effective and useful source of ultrashort wavelength radiation can then be pursued.

The authors would like to thank one of the anonymous referees for the invaluable comments and to acknowledge S. Martins for the OSIRIS ionization module and W. B. Mori, M. Marti, R. A. Fonseca, and N. Lopes for discussions. Work partially supported by FCT (Portugal) through the grants PTDC/FIS/66823/2006 and SFRH/BD/38952/2007, by the DOE grant DE-FG03-92ER40727 (UCLA), by the NSF grant PHY0936266 (UCLA), and by the European Community (project EuroLeap/028514/FP6). The simulations were performed at the expp and IST clusters in Lisbon.


\newpage

\begin{thebibliography}{99}
\bibitem{bib:joshi2} C. Joshi, Phys. Plasmas \textbf{14}, 055501 (2007).
\bibitem{bib:tatarakis} M. Tatarakis \emph{et al}., Nature (London) \textbf{415}, 280 (2002).
\bibitem{bib:langmuir} L. Tonks and I. Langmuir, Phys. Rev. \textbf{33}, 195 (1929).
\bibitem{bib:dawson} T. Tajima and J. M. Dawson, Phys. Rev. Lett. \textbf{43}, 267 (1979). 
\bibitem{bib:dreambeam} S. P. D. Mangles \emph{et al}., Nature (London) \textbf{431}, 535 (2004); C.G.R. Geddes \emph{et al}., \emph{ibid} \textbf{431}, 538 (2004); J. Faure \emph{et al}., \emph{ibid} \textbf{431}, 541 (2004). 
\bibitem{bib:leemans} W. P. Leemans \emph{et al}., Nature Phys. \textbf{2}, 696 (2006).
\bibitem{bib:tsung} F. S. Tsung \emph{et al}., Phys. Rev. Lett. \textbf{93}, 185002 (2004); W. Lu \emph{et al}., Phys. Rev. ST Accel. Beams \textbf{10}, 061301 (2007).
\bibitem{bib:bfield} S. C. Wilks, W. L. Kruer, M. Tabak, and A. B. Langdon, Phys. Rev. Lett. \textbf{69}, 1383 (1992); R. J. Mason and M. Tabak, \emph{ibid} \textbf{80}, 524 (1998); M. Borghesi \emph{et al}., \emph{ibid} \textbf{80}, 5137 (1998); A. S. Sandhu \emph{et al}., \emph{ibid} \textbf{89}, 225002 (2002). 
\bibitem{bib:bmodeexp} R. L. Savage, Jr., C. Joshi, and W. B. Mori, Phys. Rev. Lett. \textbf{68}, 946 (1992); C. H. Lai  \emph{et al}., Phys. Rev. Lett. \textbf{77}, 4764 (1996).
\bibitem{bib:bmodeth} M. Lampe, E. Ott, and J. H. Walker, Phys. Fluids \textbf{21}, 42 (1978); W. B. Mori, Phys. Rev. A \textbf{44}, 5118 (1991). 
\bibitem{bib:barut} A. O. Barut and J. P. Dowling, Phys. Rev. Lett. \textbf{68}, 3571 (1992).
\bibitem{bib:ye} P. D. Ye \emph{et al}., Phys. Rev. Lett. \textbf{74}, 3013 (1995).
\bibitem{bib:fel} J. Arthur \emph{et al}., \emph{Linac Coherent Light Source (LCLS) Conceptual Design Report} SLAC-R593 (Stanford, 2002); M. Altarelli \emph{et al}., \emph{XFEL: The European X-ray Free-Electron Laser}, technical design report, preprint DESY 2006-097 (DESY, Hamburg, 2006).
\bibitem{bib:adk} M. V. Ammosov, N. B. Delone, and V. P. Krainov, Sov. Phys. JETP \textbf{64}, 1191 (1986). 
\bibitem{bib:tochitsky} S. Ya. Tochitsky \emph{et al}., Opt. Lett. \textbf{24}, 1717 (1999).
\bibitem{bib:fonseca} R. A. Fonseca \emph{et al}., Lect. Notes Comp. Sci. \textbf{2331}, 342 (2002). 
\bibitem{bib:jackson} J. D. Jackson, \emph{Classical Electrodynamics}, 3rd Edition (Wiley, New York, 1998).
\bibitem{bib:wada} S. Wada \emph{et al}., Appl. Phys. B \textbf{57}, 435 (1993).
\bibitem{bib:joshi} C. Joshi \emph{et al}., IEEE J. Quantum Electron. \textbf{23}, 1571 (1987).
\bibitem{bib:katsouleas} T. Katsouleas, C. Joshi, and W. B. Mori, Phys. Rev. Lett. \textbf{57}, 1960 (1986).
\bibitem{bib:weibel} J. J. Su \emph{et al}., IEEE Trans. Plasma Sci. \textbf{PS-15}, 192 (1987); L. O. Silva \emph{et al}., Phys. Plasmas \textbf{9}, 2458 (2002).
\bibitem{bib:focusing} C. E. Max, J. Arons, and A. B. Langdon, Phys. Rev. Lett. \textbf{33}, 209 (1974); W. B. Mori \emph{et al}., \emph{ibid} \textbf{60}, 1298 (1988).
\bibitem{bib:raman} D. W. Forslund, J. M. Kindel, and E. L. Lindman, Phys. Fluids \textbf{18}, 1002 (1975); K. Estabrook and W. L. Kruer,  \emph{ibid} \textbf{26}, 1892 (1983); W. B. Mori \emph{et al}., Phys. Rev. Lett. \textbf{72}, 1482 (1994).
\bibitem{bib:wakes} E. Esarey \emph{et al}., IEEE Trans. Plasma Sci. \textbf{24}, 252 (1996).
\bibitem{bib:fiuza} F. Fiuza, MSc Thesis, IST/UTL (2007); F. Fiuza \emph{et al}., to be published.
\end{thebibliography}
\end{document}